\documentclass[preprint,nofootinbib]{revtex4-1}

\usepackage{amsmath}
\usepackage{amsfonts,amssymb}
\usepackage{hyperref}
\usepackage{graphicx}
\usepackage[utf8]{inputenc}

\begin{document}

\title{Cosmology and Newtonian limit in a model of gravity with nonlocally interacting metrics}

\author{Leonardo Giani$^{1,2}$}
\email{giani@thphys.uni-heidelberg.de}

\author{Tays Miranda$^{1,3}$}
%\email{taysmiranda91@gmail.com}

\author{Oliver F. Piattella$^{1, 2, 4}$}
%\email{o.piattella@thphys.uni-heidelberg.de}

\affiliation{$^1$ Physics Department, Universidade Federal do Esp\'irito Santo, Avenida Fernando Ferrari 514, 29075-910 Vit\'oria, Esp\'irito Santo, Brazil,}
\affiliation{$^2$ N\'ucleo Cosmo-ufes and PPGCosmo, Universidade Federal do Esp\'irito Santo, Avenida Fernando Ferrari 514, 29075-910 Vit\'oria, Esp\'irito Santo, Brazil,}
\affiliation{$^3$ Institute of Cosmology and Gravitation, University of Portsmouth, Dennis Sciama Building, Burnaby Road, Portsmouth, PO1 3FX, United Kingdom.}
\affiliation{$^4$ Institut f\"ur Theoretische Physik, Ruprecht-Karls-Universit\"at Heidelberg, Philosophenweg 16, 69120 Heidelberg, Germany}

\begin{abstract}
 We investigate the features of the cosmological expansion history described by a recent model of gravity characterised by two nonlocally interacting metrics. We perform a detailed analysis of the dynamical system formed by the field equations and we find no stable critical points at finite and infinite distance. Nonetheless, we show that even if the universe does not evolve towards a de Sitter attractor, the effective equation of state parameter $\omega_{\rm eff}$ always tends to $-1$, independently from the value of the free parameter $m^2$, which characterises the nonlocality of the theory. We also address the behaviour of gravity on Solar System scales and the growth of small cosmological fluctuations on small scales, in the quasi-static approximation. We find a post-Newtonian $\gamma$ parameter, a slip parameter and an effective, normalised gravitational coupling different from unity. These differences all depend on $m^2$ and are negligible if one consider the cosmological solution by which $m^2 \sim H_0^2$.  
\end{abstract}

\maketitle

\section{Introduction}

The cosmological standard model, also known as $\Lambda$CDM, succeeds in explaining a plethora of observational evidences coming from many different experiments. Such a great goal is obtained in the framework of General Relativity (GR) and with six parameters which tell us that most of the universe is made up of a cosmological constant $\Lambda$, responsible for the accelerated expansion, and cold dark matter (CDM), responsible for structure formation \cite{Ade:2015xua, Aghanim:2018eyx, Piattella:2018hvi}.

However, a cosmological constant seems to be not completely satisfactory from a fundamental point of view; the main problem is a huge discrepancy (about 56 orders of magnitude) between the observed value of $\Lambda$ and the predicted value for the vacuum expectation value computed in quantum field theory, demanding thus a huge fine-tuning of a bare cosmological constant \cite{Weinberg:1988cp, Martin:2012bt}. In order to overcome this problem, alternatives to $\Lambda$ are explored and generally dubbed as dark energy (DE) referring both to more exotic matter components (e.g. quintessence, $k$-essence) and to geometrical effects in theories of gravity which extend or modify GR. See e.g. Ref.~\cite{Amendola:2015ksp}.

Since GR is very successful on a large window of scales, the problem of DE is usually addressed geometrically by adding some infrared modifications in the Einstein Hilbert action, such that these new terms become effective only on cosmological scales. Among those possible modifications, an interesting class of models arises when we relax the request for our gravitational theory to be local. Without this paradigm, it is possible to introduce nonlocal deformations to the Einstein-Hilbert action which can reproduce effectively a cosmological constant. A very interesting class of nonlocal models makes use of terms which are proportional to some negative power of a differential operator acting on the Ricci scalar, like the Deser-Woodard \cite{Deser:2007jk} and Maggiore-Mancarella models \cite{Maggiore:2014sia}, and are able to reproduce a cosmological history similar to the one of the $\Lambda$CDM model. See also Refs.~\cite{Koivisto:2008xfa, Koivisto:2008dh, Park:2012cp, Deser:2013uya, Woodard:2014iga, Dirian:2014ara, Barreira:2014kra, Dirian:2014bma, Dirian:2016puz, Nersisyan:2016hjh, Maggiore:2016gpx, Nersisyan:2017mgj, Park:2017zls,  Belgacem:2017cqo, Amendola:2019fhc} for an incomplete of investigations of some nonlocal models, their cosmological constraints and also their stability against ghosts.

In this work we investigate the nonlocal gravity model proposed in Ref.~\cite{Vardanyan:2017kal}, and characterised by a bimetric theory of gravity with a nonlocal interaction term depending on the Ricci scalars of the two metrics. As it is showed in Ref.~\cite{Vardanyan:2017kal}, already the simplest version of the model allows to reproduce a viable cosmological history with no appearance of ghosts. However, a complete analysis of the dynamics of the model is still missing. In this paper we fill this gap by performing a dynamical system analysis of the simplest version of the theory in a cosmological setting. We find no stable critical points, neither at finite or infinite distance, and a hyperplane of unstable critical points at infinite distance. 

The absence of a de Sitter stable attractor might suggest that any accelerated phase of expansion similar to the one of the $\Lambda$CDM model is transitory. In order to establish if this is the case, we analyse the asymptotic behaviour of the effective equation of state for this model and prove that it always tends to $-1$, thereby mimicking in this sense the $\Lambda$CDM model, even if the Hubble parameter does not tend to a constant.

Finally, we also consider the Newtonian limit of the model and small cosmological fluctuations on small scales, in order to understand how nonlocality manifests itself at Solar System scales and how it affects the growth of small dust homogeneities. We find modifications of the post-Newtonian parameter $\gamma$, of the slip parameter and of the effective gravitational coupling with respect to the GR case. However, these modifications are not dangerous for the viability of the model. 

The paper is structured as follows. In Sec.~\ref{Sec:nonlocalmodel} we introduce the model and its field equations and set up the dynamical system for the cosmological scenario, looking for its critical points and their stability. In Sec.~\ref{Sec:Qualitativeanalysis} we present a qualitative analysis of the field equations, corroborating the dynamical system one and proving that the effective equation of state always tends to $-1$. In Sec. ~\ref{Sec:Newtonian Limit} we considered first order scalar cosmological perturbations, the Newtonian limit and the growth of small dust fluctuations in the quasi-static regime. Finally, in Sec.~\ref{Sec:Concl} we discuss our results and present our conclusions. Units $c = 1$ are used throughout the paper.

\section{The nonlocal model}\label{Sec:nonlocalmodel}

In the model proposed in Ref.~\cite{Vardanyan:2017kal}, hereafter referred to as VAAS (from the initials of the authors' surnames), nonlocality intervenes in the interaction between two metrics, the physical one $g_{\mu\nu}$, to which matter couples and by which geodesics are defined, and an auxiliary one $f_{\mu\nu}$. The action is the following:
\begin{eqnarray}
	S = \frac{M_{\rm Pl}^2}{2}\int d^4x\sqrt{-g}R + \frac{M_f^2}{2}\int d^4x\sqrt{-f}R_f\nonumber\\ -\frac{M_{\rm Pl}^2}{2}\int d^4x\sqrt{-g}\alpha\left(R_f\frac{1}{\Box}R + R\frac{1}{\Box}R_f\right) + S_{\rm matter}[g,\Psi]\;, 
\end{eqnarray}
where $\alpha$ is a coupling constant tuning the nonlocal interaction, $R$ is the Ricci scalar corresponding to $g_{\mu\nu}$, $R_f$ is the Ricci scalar corresponding to $f_{\mu\nu}$, and $\Psi$ is a shortcut notation for all the matter fields, including CDM. It turns out, from computing the Bianchi constraints, that $R_f$ must be constant and the field equations obtained upon variation of the action with respect to $g_{\mu\nu}$ can be cast as follows:
\begin{eqnarray}\label{FieldequationsNL}   
(1 - 2\alpha V)G_{\mu\nu} + m^2(1 - U/2)g_{\mu\nu} + 2\alpha\nabla_\mu\nabla_\nu V + \alpha \nabla^\rho V\nabla_\rho Ug_{\mu\nu}\nonumber\\ - 2\alpha\nabla_{(\mu}U\nabla_{\nu)}V = \frac{1}{M_{\rm Pl}^2}T_{\mu\nu}\;,
\end{eqnarray}
where $G_{\mu\nu}$ is the Einstein tensor computed from the physical metric, $\nabla_\mu$ is the covariant derivative computed with the standard Levi-Civita connection defined from $g_{\mu\nu}$, and $m^2 \equiv -2\alpha R_f$. The notation $(\mu\nu)$ indicates symmetrisation, i.e. explicitly:
\begin{equation}
	\nabla_{(\mu}U\nabla_{\nu)}V \equiv \frac{1}{2}\left(\nabla_{\mu}U\nabla_{\nu}V + \nabla_{\nu}U\nabla_{\mu}V\right)\;.
\end{equation} 
The two auxiliary fields $U$ and $V$ are introduced in order to localise the theory, and therefore to make it more easily treatable, and they satisfy the following equations:
\begin{equation}\label{auxiliaryeqsNL}
	\square U = R\;, \qquad \square V = R_f = -\frac{m^2}{2\alpha}\;.
\end{equation}
Let us now investigate the cosmological evolution realised in the above model and to this purpose let us assume a Friedmann-Lema\^itre-Robertson-Walker (FLRW) line element:
\begin{equation}
	ds^2 = g_{\mu\nu}dx^\mu dx^\nu = -dt^2 + a(t)^2\delta_{ij}dx^idx^j\;,
\end{equation}
with flat spatial hypersurfaces. Along with this choice, we also assume that $U$ and $V$ are functions of time only. The field equations \eqref{FieldequationsNL} then become:
\begin{eqnarray}
\label{modFriedeq}	(1 - 2\alpha V)3H^2 - m^2(1 - U/2) + 2\alpha\ddot V - \alpha\dot U\dot V = \frac{\rho}{M_{\rm Pl}^2}\;,\\
\label{modacceq}	-(1 - 2\alpha V)g_{ij}(2\dot H + 3H^2) + m^2(1 - U/2)g_{ij} - 2\alpha H\dot Vg_{ij}\nonumber\\ - \alpha\dot U\dot Vg_{ij} = \frac{1}{M_{\rm Pl}^2}Pg_{ij}\;,
\end{eqnarray}
where the dot denotes derivation with respect to the cosmic time $t$. Equation \eqref{modFriedeq} is the $\mu = \nu = 0$ equation, i.e. the modified Friedmann equation, whereas Eq.~\eqref{modacceq} is the $\mu = i$, $\nu = j$ equation, i.e. the modified acceleration equation. The equation for $\mu = 0$ and $\nu = i$ is identically vanishing. We have assumed a matter content given by a perfect fluid:
\begin{equation}
	T_{\mu\nu} = (\rho + P)u_\mu u_\nu + Pg_{\mu\nu}\;.
\end{equation}
The auxiliary equations \eqref{auxiliaryeqsNL}, after working out the box operator, become:
\begin{eqnarray}
\label{UandVeqs}	\ddot U + 3H\dot U = -R = -6(\dot H + 2H^2)\;, \qquad \ddot V + 3H\dot V = \frac{m^2}{2\alpha}\;,
\end{eqnarray}
Combining Eqs.~\eqref{modFriedeq} and \eqref{modacceq} and defining $\dot{U} \equiv X$ and $\dot{V} \equiv Y$ in Eq.~\eqref{UandVeqs}, one gets:
\begin{subequations}\label{dsfinitedistance2}
\begin{eqnarray}
    \dot{H} &=&\frac{1}{1 - 2 \alpha V} \left[\frac{\rho - P}{2 M_{\rm Pl}^{2}} + \frac{m^{2}}{2}\left(1 - U \right) + 2\alpha H Y\right] - 3 H^{2}\;,\\
    \dot{\rho} &=& -3 H(\rho + P)\;,\\
    \dot{X} &=& - 3 H X - 6 H^{2} - \frac{6}{1 - 2 \alpha V} \left[\frac{\rho - P}{2 M_{\rm Pl}^{2}} + \frac{m^{2}}{2}\left(1 - U \right) + 2\alpha H Y\right]\;,\\
    \dot{Y} &=& \frac{m^{2}}{2 \alpha} - 3 H Y \;,\\
    \dot{U} &=& X\;,\\
    \dot{V} &=& Y\;.
\end{eqnarray}
\end{subequations}
This is the dynamical system which desrcibes the expansion history in the VAAS model. Recall that the modified Friedmann equation provides a constraint, hence we must also take into account Eq.~\eqref{modFriedeq}:
\begin{equation}\label{modFriedeq2}
	(1 - 2\alpha V)3H^2 + \frac{m^2U}{2} - 6H\alpha Y - \alpha XY = \frac{\rho}{M_{\rm Pl}^2}\;.
\end{equation}

\subsection{Critical points at finite distance}

A critical point of the dynamical system \eqref{dsfinitedistance2} is a point in the phase space for which the right hand sides of the equations \eqref{dsfinitedistance2} vanish.

Consider first the case where $2\alpha V = 1$, which is special because it eliminates $\dot H$ from the field equations. From Eq.~\eqref{UandVeqs} we can see that $V = 1/(2\alpha) =$ constant only if $m^2 = 0$. The latter is a parameter of the theory so it is not necessarily vanishing and thus, unless $m^2 = 0$, $V$ cannot be constant. With $V = 1/(2\alpha) =$ constant, then $\dot V = Y = 0$ and thus from Eq.~\eqref{modFriedeq2} we get $\rho = 0$. From Eq.~\eqref{modacceq} we also get $P = 0$. We are just left with Eq.~\eqref{UandVeqs} for $U$ and $H$ is completely arbitrary, since we have lost the equation ruling its dynamics. Since providing a suitable $H$ is one of the objectives of the model, we do not consider this possibility anymore.

Now we consider $2\alpha V \neq 1$. The right hand sides of the last three equations of the dynamical system~\eqref{dsfinitedistance2} vanish when $X = Y = 0$ and $m^2 = 0$. Again, the latter is a parameter of the theory so it is not necessarily vanishing. Thus if $m^2 \neq 0$ we can already conclude that there are no critical points at finite distance. 

Let us consider now the subclass of theories for which $m^2 = 0$, i.e. with $R_f = 0$. Demanding the vanishing of the right hand sides of the first three equations of system~\eqref{dsfinitedistance2} and from Eq.~\eqref{modFriedeq2} we have, taking into account $X = Y = m^2 = 0$:
\begin{eqnarray}
	3H^2(1 - 2 \alpha V) = \frac{\rho - P}{2 M_{\rm Pl}^{2}}\;,\\
	-3 H(\rho + P) = 0\;,\\
	H^2(1 - 2\alpha V) = \frac{P - \rho}{2 M_{\rm Pl}^{2}}\;,\\
	3H^2(1 - 2\alpha V) = \frac{\rho}{M_{\rm Pl}^{2}}\;.
\end{eqnarray}
Now, from the second equation above we either have that $H = 0$ or $P = -\rho$, i.e. a vacuum energy equation of state is required. If our fluid model has not such equation of state, then the only possibility is $H = 0$ and thus $\rho = P = 0$. This critical point represents Minkowski space. Note that $U$ and $V$ may assume whatever constant value, except $V = 1/(2\alpha)$.

On the other hand, let us assume that indeed the fluid content satisfies a vacuum energy equation of state, i.e. $P = -\rho$. In this case, the above system becomes:
\begin{eqnarray}
	3H^2(1 - 2 \alpha V) = \frac{\rho}{M_{\rm Pl}^{2}}\;,\\
	H^2(1 - 2\alpha V) = -\frac{\rho}{M_{\rm Pl}^{2}}\;.
\end{eqnarray}
Summing the two equations we arrive at:
\begin{equation}
	4H^2(1 - 2 \alpha V) = 0\;.
\end{equation}
Since $2\alpha V \neq 1$, we have again $H = 0$ and thus the same Minkowski critical point as before. There is a caveat here. When $m^2 = 0$ the evolution of $U$ is disentangled from the one of the other variables and the dynamical system~\eqref{dsfinitedistance2} can be reduced to:
\begin{subequations}\label{dsm2zero}
\begin{eqnarray}
    \dot{H} &=&\frac{1}{1 - 2 \alpha V} \left[\frac{\rho - P}{2 M_{\rm Pl}^{2}} + 2\alpha H Y\right] - 3 H^{2}\;,\\
    \dot{\rho} &=& -3 H(\rho + P)\;,\\
    \dot{Y} &=& - 3 H Y \;,\\
    \dot{V} &=& Y\;.
\end{eqnarray}
\end{subequations}
Note that $U$ drops out also from the constraint \eqref{modFriedeq2} for $m^2 = 0$, so in this sense we are entitled to claim that its evolution is disentangled from the one of the other variables.

From system \eqref{dsm2zero} is not difficult to see that there is a critical point 
\begin{equation}
	P = -\rho = -M^2_{\rm Pl}\left(1-2\alpha V\right)3H^2 =\mbox{ constant}\;,
\end{equation}
which represents a de Sitter phase. With $H > 0$ constant, the equation for $U$ becomes:
\begin{equation}
	\ddot U + 3H\dot U = -12H^2\;,
\end{equation}
which has the special, non-constant solution $U = -4Ht$.

We can then conclude the first part of our analysis as follows:
\begin{itemize}
	\item For $m^2 \neq 0$ there are no critical points at finite distance;
	\item For $m^2 = 0$ and $U =$ constant, the only critical point at finite distance represents Minkowski space.
	\item For $m^2 = 0$, $U = -4Ht$ and $H$ constant, the only critical point at finite distance represents a de Sitter space.
\end{itemize}

\subsection{Critical points at infinite distance} 

In order to investigate the critical points at infinity, we build the Poincaré hypersphere in the phase space augmented of one dimension. See Ref.~\cite{sansone2016non} for the mathematical details of the construction. The equation of the hypersphere is the following: 
\begin{eqnarray}\label{sphere}
    h^{2} + r^{2} + p^{2} + x^{2} + y^{2} + u^{2} + v^{2} + z^{2} = 1\;,
\end{eqnarray}
where we have defined:
\begin{eqnarray}
    H \equiv \frac{h}{z}\;, \quad \rho \equiv \frac{r}{z}\;, \quad P \equiv \frac{p}{z}\;, \quad X \equiv \frac{x}{z}\;, \quad Y \equiv \frac{y}{z}\;, \quad U \equiv \frac{u}{z}\; \quad \textrm{and} \quad V \equiv \frac{v}{z}\;.
\end{eqnarray}
Since we assume a barotropic equation of state:
\begin{equation}
	P = (\gamma - 1)\rho\;,
\end{equation}
the pressure is no more to be considered as an independent phase space variable and therefore we drop it from the construction of the Poincar\'e sphere and the dynamical system \eqref{dsfinitedistance2} thus becomes:
\begin{subequations}\label{dsinfinitedistance1}
\begin{eqnarray}
    z\dot{h} &=& A (1 - h^{2}) -  h \left(r B + x C + y D + u E + v F \right)\;,\\
    z\dot{r} &=& B (1 - r^{2})-  r \left(h A + x C + y D + u E + v F \right)\;,\\
    z\dot{x} &=& C (1 - x^{2}) - x \left(h A + r B + y D + u E + v F \right)\;,\\
    z\dot{y} &=& D (1 - y^{2}) - y \left(h A + r B + x C + u E + v F \right)\;,\\
    z\dot{u} &=& E (1 - u^{2}) - u \left(h A + r B + x C + y D + v F \right)\;,\\
    z\dot{v} &=& F (1 - v^{2}) - v \left(h A + r B + x C + y D + u E \right)\;,\\
    z\dot{z} &=& -\left(h A + r B + x C + y D + u E + v F \right)\;,
\end{eqnarray}
\end{subequations}
where the equation for $\dot{z}$ is obtained from Eq.~\eqref{sphere} and the terms $A$, $B$, $C$, $D$, $E$ and $F$ are defined as follows:
\begin{subequations}
\begin{eqnarray}
    A &\equiv & \frac{z}{z - 2 \alpha v} \left[\frac{z(r - p)}{2M_{\rm Pl}^{2}} + \frac{m^{2}z}{2}\left(z - u\right) + 2\alpha h y\right] - 3h^2\;,\\
    B &\equiv & - 3 h (r + p)\;,\\
    C &\equiv & - 3 h x - 6 h^{2} -\frac{6 z}{z - 2 \alpha v} \left[\frac{z(r - p)}{2M_{\rm Pl}^{2}} + \frac{m^{2}z}{2}\left(z - u\right) + 2\alpha h y\right]\;,\\
    D &\equiv & \frac{m^{2}z^2}{2 \alpha} - 3 h y\;,\\
    E &\equiv & zx\;,\\
    F &\equiv & zy\;.
\end{eqnarray}
\end{subequations}
The critical points of the above system corresponding to $z = 0$ are critical points at infinity. In order to find them, let us first define a new time parameter such that $\frac{z}{3}\frac{d}{dt} \equiv \frac{d}{d \tau} \equiv\;'$ and let us also define the following functions:
\begin{eqnarray}
    G \equiv h \left(h^{2} + \gamma r^{2} + x^{2} + y^{2} + 2 h x -1\right)\;,\\
    \tilde G \equiv h \left(h^{2} + \gamma r^{2} + x^{2} + y^{2} + 2 h x - \gamma\right)\;,
\end{eqnarray}
The dynamical system \eqref{dsinfinitedistance1} for $z =0$ can thus be written as:
\begin{subequations}\label{dsinfinitedistance2}
\begin{eqnarray}
    h' &=& h G\;, \label{dsinfinitedistance2h}\\
    r' &=& r \tilde G\;,\\
    x' &=& x G - 2 h^{2}\;,\label{dsinfinitedistance2x}\\
    y' &=& y G\;, \label{dsinfinitedistance2y}\\
    u' &=& u\left(G + h\right)\;,\\
    v' &=& v\left(G + h\right)\;.
\end{eqnarray}
\end{subequations}
It is important to emphasise that the solutions of the above system must be compatible with the Friedmann equation, which in terms of the variables on the Poincaré sphere provides:
\begin{eqnarray}\label{Friedmanninfinitedistance}
    \left(1 - 2\alpha \frac{v}{z}\right)\frac{h^{2}}{z^{2}} + \frac{m^{2} u}{6 z} - \frac{\alpha y}{3 z^{2}} (6h + x) - \frac{r}{3 z M_{\rm Pl}^{2}} = 0\;.
\end{eqnarray}
Multiplying the above equation for $z^{3}$ and then considering $z = 0$, we obtain:
\begin{eqnarray}
    \alpha v h^{2} = 0.
\end{eqnarray}
Therefore, at infinity Friedmann equation imposes that at least one among $\alpha$, $v$, $h$ is vanishing. It is easy to see that when $h = 0$ the function $G$ vanishes identically, so we have a critical hyperplane in the variables space corresponding to Minkowski spacetime. 

We discuss later the stability of this critical hypersurface, and focus now on the only other interesting case,\footnote{We do not consider $\alpha = 0$ since it simply turns off the nonlocal interacting term.} $v = 0$. In this instance, Friedmann equation becomes:
\begin{equation}\label{Friedeqvzero}
    h^2  -\alpha y \left(2h + \frac{x}{3} \right) = 0 \;. 
\end{equation}
Since $h \neq 0$, then from Eqs.~\eqref{dsinfinitedistance2h} and \eqref{dsinfinitedistance2y} we have that:
\begin{equation}
	y = h + K\;,
\end{equation}
where $K$ is an integration constant. From Eq.~\eqref{Friedeqvzero} we then have:
\begin{equation}
	x = \frac{3h^2}{\alpha y} - 6h = \frac{3h^2}{\alpha(h + K)} - 6h\;.
\end{equation}
This result allows us to rewrite Eq.~\eqref{dsinfinitedistance2x} as follows:
\begin{equation}\label{xhequation}
    x' = G\left[\frac{3h^2}{\alpha \left(h + K\right)} - 6h\right] - 2h^2 \; ,
\end{equation}
and we see that it is impossible to have both \eqref{dsinfinitedistance2h} and \eqref{xhequation} vanishing  without $G=h=0$, and so there are no other critical points, but $h = 0$, at infinite distance.

\subsection{Linearisation and stability of the critical point}

In order to investigate the stability of the critical points at $h = 0$, we linearise system \eqref{dsinfinitedistance2} around the critical point $h_{0} =0$:
\begin{equation}
	h = h_{0} + \epsilon\;,\quad r = r_{0} + \eta\;,\quad x = x_{0} + \chi\;,\quad y = y_{0} + \varphi\;,\quad u = u_{0} + \lambda\;,\quad v = v_{0} + \sigma\;.
\end{equation} 
Consequently, the linearised dynamical system is given by:
\begin{subequations}\label{linearizedds}
\begin{eqnarray}
    \epsilon ' &=& 0 \; ,\label{linearizeddsepsilon}\\
    \eta ' &=& r_{0} \left(\gamma r_{0}^{2} + x_{0}^{2} + y_{0}^{2} - \gamma \right) \epsilon \; ,\\
    \chi ' &=& x_{0} \left(\gamma r_{0}^{2} + x_{0}^{2} + y_{0}^{2} - 1 \right) \epsilon \; ,\\
    \varphi ' &=& y_{0} \left(\gamma r_{0}^{2} + x_{0}^{2} + y_{0}^{2} - 1 \right) \epsilon \; ,\\
    \sigma ' &=& u_{0} \left(\gamma r_{0}^{2} + x_{0}^{2} + y_{0}^{2}\right) \epsilon \; ,\\
    \lambda ' &=& v_{0} \left(\gamma r_{0}^{2} + x_{0}^{2} + y_{0}^{2}\right) \epsilon \; .
\end{eqnarray}
\end{subequations}
Usually the stability of the critical point is studied by means of the Jacobian matrix, but unfortunately it is degenerate in our present case. However, we easily recognise that in the above system of equations the perturbation $\epsilon$ is constrained to be a constant by \eqref{linearizeddsepsilon}. Note also that the combinations $\left(\gamma r_{0}^{2} + x_{0}^{2} + y_{0}^{2} - 1 \right)$ and $\left(\gamma r_{0}^{2} + x_{0}^{2} + y_{0}^{2} - \gamma \right)$ are constant as long as we assume a time-independent equation of state.  Thus all the perturbations with the exception of $\epsilon$, which is constant, grow linearly with time displaying thus an unstable behaviour.

We can conclude this section with the first important result of the paper: a cosmological model based on the VAAS theory has no stable critical point. Minkowski space is a critical point at infinity, but it is an unstable one. In particular, it cannot be an attractor and therefore we do not expect a cosmological evolution for which $H \to 0$ in the future, as it is the case for standard cosmology in the absence of a cosmological constant. We discuss this peculiarity in more detail in the next Section. 

\section{Qualitative analysis of the cosmological evolution}\label{Sec:Qualitativeanalysis}

In this section we investigate in detail some characteristics of the cosmological evolution in the VAAS theory. To this purpose, let us write down Eqs.~\eqref{modFriedeq}, \eqref{modacceq} and \eqref{UandVeqs} here, using the e-folds number $N \equiv \ln a$ as independent variable. Moreover, it is convenient to use as variables:
\begin{equation}
	\tilde V \equiv 1 - 2\alpha V\;, \qquad \xi \equiv \frac{H'}{H}\;,
\end{equation}
where the prime denotes hereafter derivation with respect to $N$ (it has nothing to do with the new time parameter $\tau$ introduced in the previous section). Our nonlocal cosmological evolution is described by the following system of equations:
\begin{subequations}\label{systemVAAS}
\begin{eqnarray}
\label{Friedeq} 3\tilde V + \frac{m^2U}{2H^2} + 3\tilde V' + \frac{U'\tilde V'}{2} = \frac{\rho}{M_{\rm Pl}^2H^2}\;,\\
\label{acceq} -\tilde V\left(3 + 2\xi\right) + \frac{m^2}{H^2}(1 - U/2) + \tilde V' + \frac{U'\tilde V'}{2} = \frac{1}{M_{\rm Pl}^2H^2}P\;,\\
U'' + \left(3 + \xi\right)U' + 6\left(2 + \xi\right) = 0\;,\\ 
\tilde V'' + \left(3 + \xi\right)\tilde V' = -\frac{m^2}{H^2}\;.
\end{eqnarray}
\end{subequations}
As a first step, we numerically solve the above system \eqref{systemVAAS} in order to understand the main features of the cosmological evolution that we want to explain later analytically based on some approximations. We start our numerical integration from $N_i = -15$, where we assume $U_i = V_i = 0$ (hence $\tilde V_i = 1$) along with their derivatives, i.e. $U'_i = V'_i = \tilde V_i' = 0$, and the Hubble factor normalised to the Hubble constant is given at $N_i$ as follows:
\begin{equation}
	h_i \equiv \frac{H_i}{H_0} = \Omega_{\rm m0}e^{-3N_i} + \Omega_{\rm r0}e^{-4N_i}\;,
\end{equation}
with $\Omega_{\rm m0} = 0.31$ and $\Omega_{\rm r0} = 9.2\times 10^{-5}$. This $h_i$ (which has nothing to do with the $h$ introduced in the previous section for the investigation of the critical points at infinity) is the same as the one in standard cosmology because we do not want to spoil the early-times history of the universe, in particular the thermal history. Indeed, one can see from Eqs.~\eqref{Friedeq} and \eqref{acceq}, that the chosen initial conditions guarantee that the standard Friedmann equations hold at early-times.

\begin{figure}
	\includegraphics[width=0.5\columnwidth]{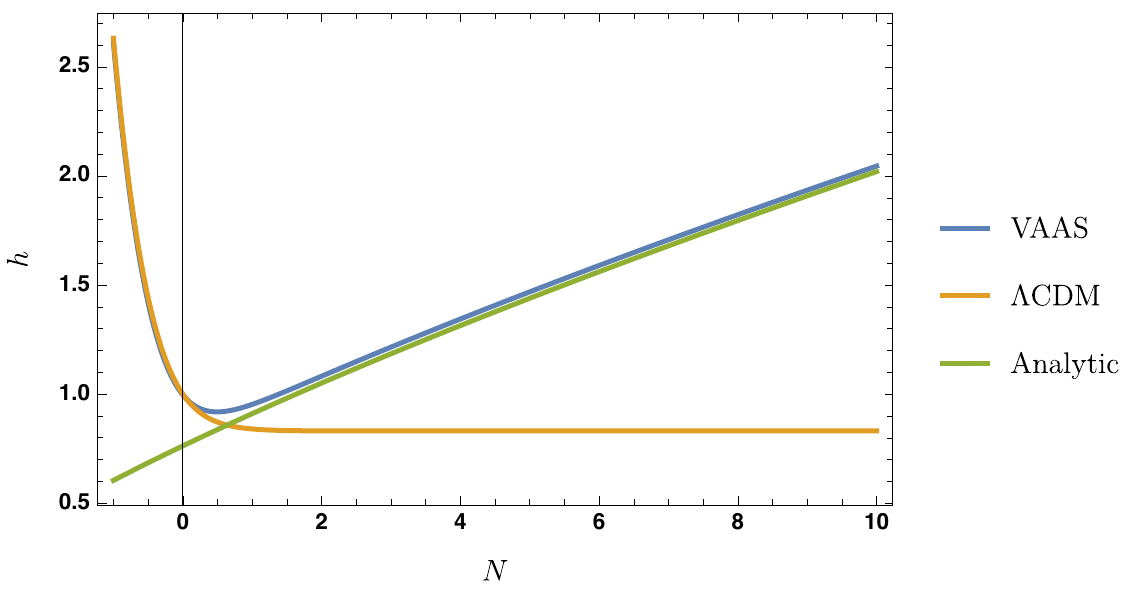}\includegraphics[width=0.5\columnwidth]{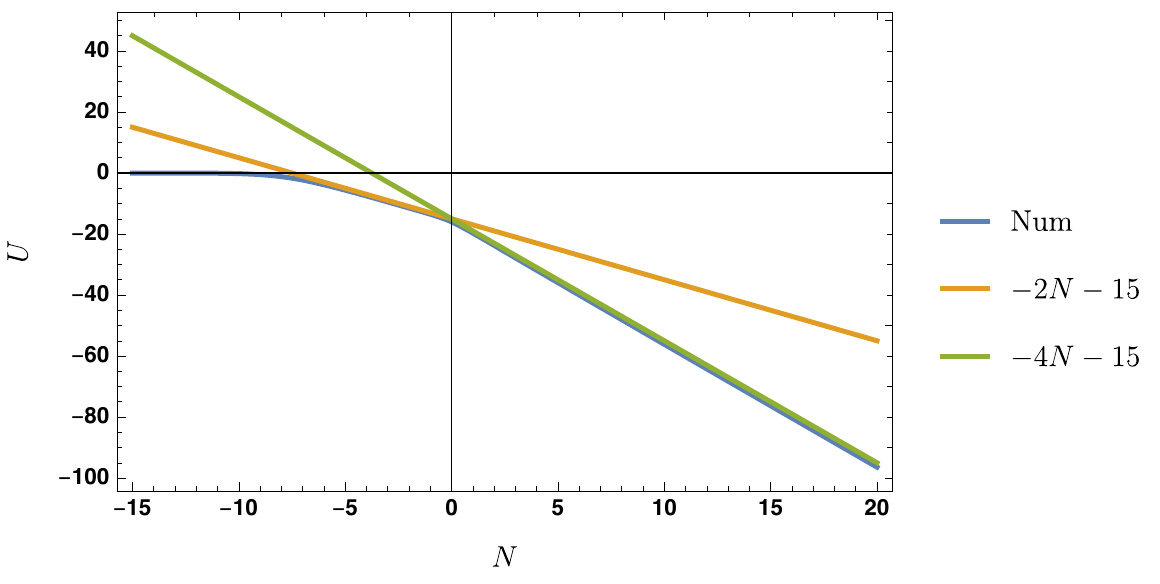}
	\caption{Evolutions of $h$ and $U$ for the model of Ref.~\cite{Vardanyan:2017kal} (VAAS) and for the $\Lambda$CDM, along with their analytic approximations ($h_{\rm an} = (4N + 15)^{3/4}/10$) which are computed in the text, for the choice $m^2/H_0^2 = 0.232$, the same of Ref.~\cite{Vardanyan:2017kal}.}
	\label{Fig:hU}
\end{figure}

\begin{figure}
	\includegraphics[width=0.5\columnwidth]{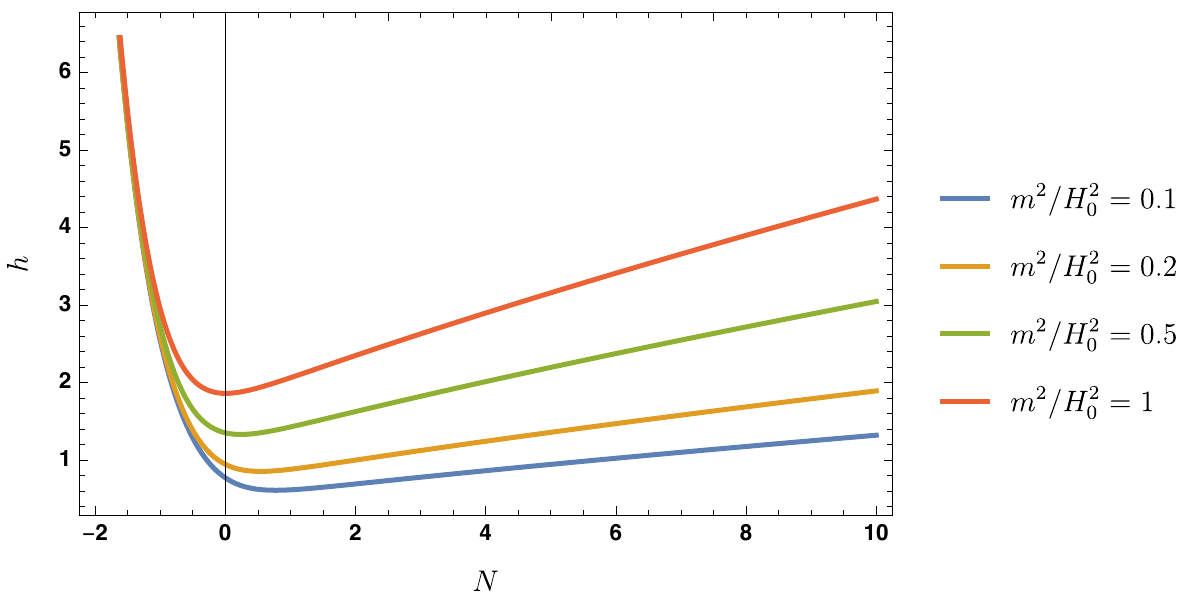}\includegraphics[width=0.5\columnwidth]{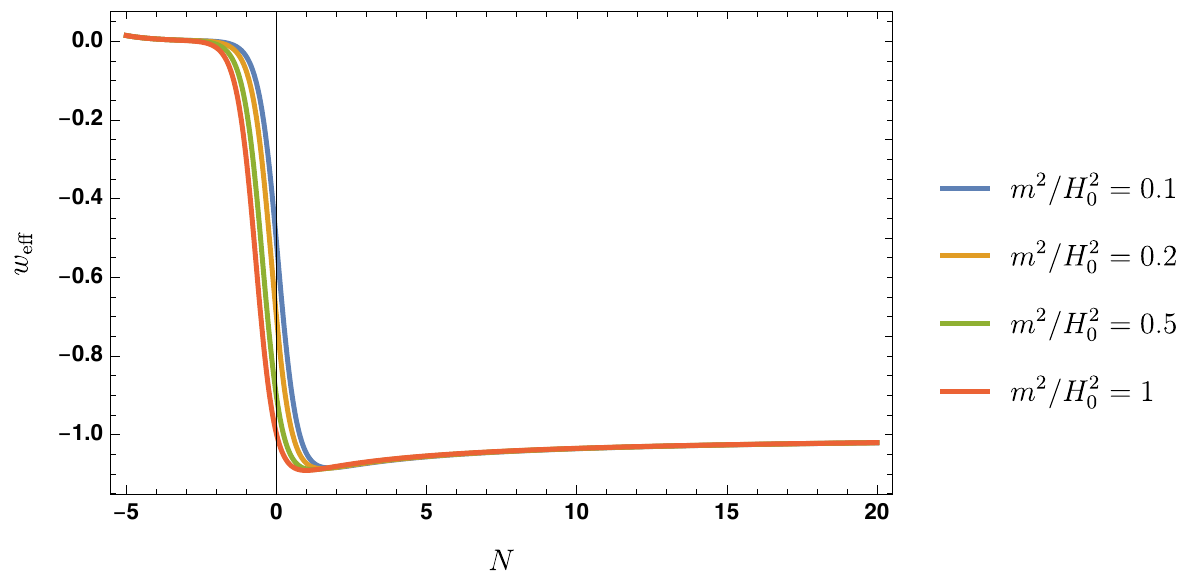}
	\caption{Evolutions of $h$ and $w_{\rm eff} \equiv -1 - 2\xi/3$ for the choices of $m^2/H_0^2$ shown in the legend.}
	\label{Fig:hweff}
\end{figure}

\begin{figure}
	\includegraphics[width=0.5\columnwidth]{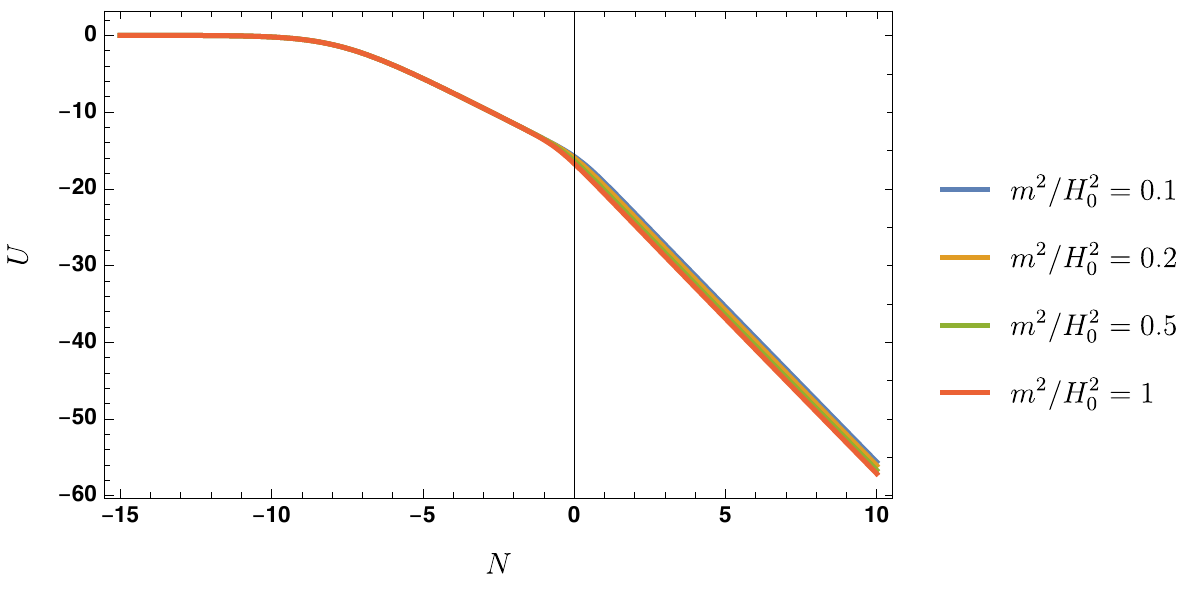}\includegraphics[width=0.5\columnwidth]{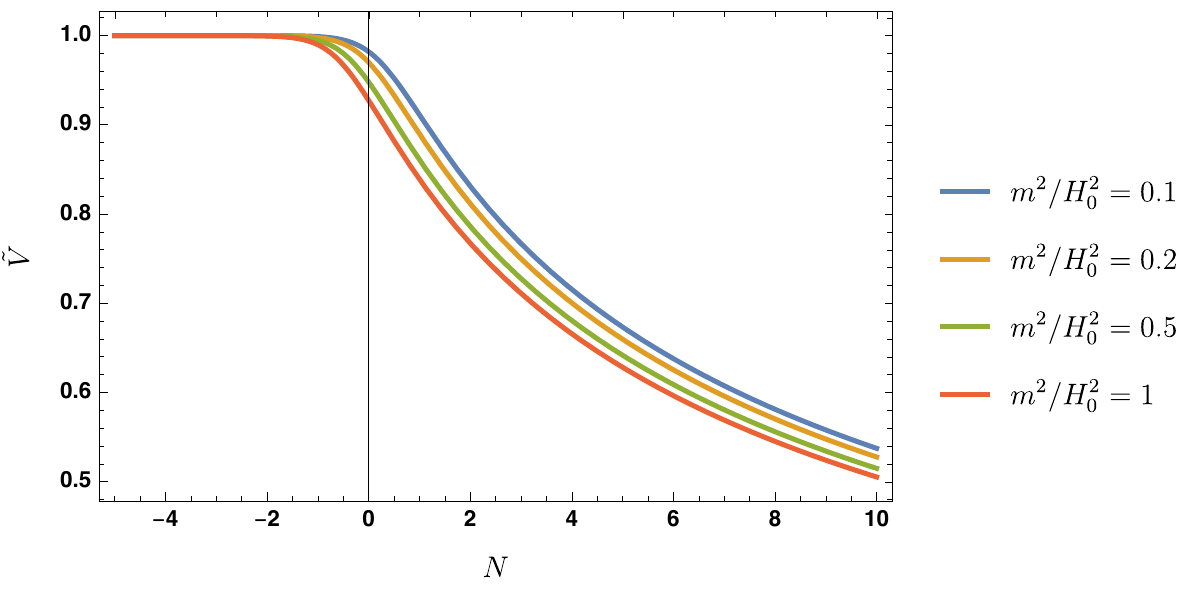}
	\caption{Evolutions of $U$ and $\tilde V$ for the choices of $m^2/H_0^2$ shown in the legend.}
	\label{Fig:UV}
\end{figure}

From Figs.~\ref{Fig:hU}, \ref{Fig:hweff} and \ref{Fig:UV} one sees that $h(N)$ decreases, as expected in standard cosmology, but then it increases, displaying thus a minimum. This behaviour seems to be independent from the value of $m^2$, which only tunes the value of $N$ corresponding to the minimum. Moreover, $h$ grows faster than $h'$ and therefore the effective equation of state, which is defined as:
\begin{equation}
	w_{\rm eff} \equiv -1 - \frac{2}{3}\xi\;,
\end{equation} 
seems to tend to $-1$, regardless the value of $m^2$. The behaviours of $U$ and $V$ are also qualitatively independent from $m^2$ and $\tilde V$ seems to be always positive. We try now to capture these qualitative features through an approximate, analytic analysis.  

Calling $X \equiv U'$ and $Y \equiv \tilde V'$, the equations:
\begin{eqnarray}
 X' + \left(3 + \xi\right)X + 6\left(2 + \xi\right) = 0\;,\\ 
 Y' + \left(3 + \xi\right)Y = -\frac{m^2}{H^2}\;,
\end{eqnarray} 
have formal solutions:
\begin{eqnarray}
	X(N) = C_1e^{-F(N)} - 6e^{-F(N)}\int^N_{N_i} d\bar Ne^{F(\bar N)}[2 + \xi(\bar N)]\;,\\
	Y(N) = C_2e^{-F(N)} - e^{-F(N)}\int^N_{N_i} d\bar Ne^{F(\bar N)}\frac{m^2}{H^2(\bar N)}\;,
\end{eqnarray}
with $C_1$ and $C_2$ integration constants and:
\begin{equation}
	F(N) \equiv \int^N_{N_i}d\bar N[3 + \xi(\bar N)]\;.
\end{equation}
Note that:
\begin{equation}
	C_1 = X(N_i)\;, \qquad C_2 = Y(N_i)\;,
\end{equation} 
and we have fixed them to be vanishing. Hence:
\begin{eqnarray}
\label{Xsol}	X(N) = -6e^{-F(N)}\int^N_{N_i} d\bar Ne^{F(\bar N)}[2 + \xi(\bar N)]\;,\\
	Y(N) = -e^{-F(N)}\int^N_{N_i} d\bar Ne^{F(\bar N)}\frac{m^2}{H^2(\bar N)}\;.
\end{eqnarray}
If we make the assumption $\xi + 2 > 0$, then we see that $X$ and $Y$ are always negative, i.e. $U$ and $\tilde V$ always decrease. In particular, since its initial value is vanishing, then $U$ is always negative. We can also prove that:
\begin{equation}
	X + 6 > 0\;,
\end{equation} 
which shall be very useful later. Indeed, rewrite the solution for $X$ as follows:
\begin{equation}
	X(N) = -6e^{-F(N)}\int^N_{N_i} d\bar Ne^{F(\bar N)}[3 + \xi(\bar N)] + 6e^{-F(N)}\int^N_{N_i}d\bar Ne^{F(\bar N)}\;.
\end{equation}
The first integral can be cast as:
\begin{equation}
	X(N) = -6e^{-F(N)}\int^N_{N_i} d\bar N\frac{d(e^{F(\bar N)})}{d\bar N} + 6e^{-F(N)}\int^N_{N_i}d\bar Ne^{F(\bar N)}\;,
\end{equation}
and thus:
\begin{equation}
	X(N) = -6 + 6e^{-F(N)} + 6e^{-F(N)}\int^N_{N_i}d\bar Ne^{F(\bar N)}\;.
\end{equation}
Being the second and third terms on the right hand side strictly positive, we have then that $X(N) > -6$.

The assumption $\xi + 2 > 0$ is reasonable because we want in the past a radiation-dominated epoch, for which $\xi = -2$, then a matter-dominated one, for which $\xi = -3/2$, and then an accelerated expansion, for which $\xi > -1$ (this comes from demanding $\ddot a > 0$).

Let us address now the evolution of the localised fields during the former two phases. During the radiation-dominated epoch one has $\xi = -2$ and thus:
\begin{equation}
	X = 0\;,
\end{equation}
This implies that $U$ is a constant, and this constant must be zero, because of our initial condition. This is in qualitative agreement with Fig.~\ref{Fig:hU}. On the other hand,
\begin{equation}
	Y(N) = -e^{-(N-N_i)}\int^N_{N_i} d\bar Ne^{(\bar N-N_i)}\frac{m^2}{H_0^2\Omega_{\rm r0}e^{-4\bar N}} = -\frac{m^2}{5H_0^2\Omega_{\rm r0}}e^{-N}\left(e^{5N} - e^{5N_i}\right)\;,
\end{equation}
from which:
\begin{equation}
	\tilde V = -\frac{m^2}{20H_0^2\Omega_{\rm r0}}e^{4N} - \frac{m^2}{5H_0^2\Omega_{\rm r0}}e^{5N_i - N} + C_3\;.
\end{equation}
Since $\tilde V(N_i) = 1$, we then have:
\begin{equation}
	\tilde V = -\frac{m^2}{20H_0^2\Omega_{\rm r0}}e^{4N} - \frac{m^2}{5H_0^2\Omega_{\rm r0}}e^{5N_i - N} + \frac{m^2}{4H_0^2\Omega_{\rm r0}}e^{4N_i} + 1\;.
\end{equation}
This is very small for $N$ large and negative, so one can basically take $\tilde V = 1$.

During the matter-dominated epoch we have $\xi = -3/2$ and thus:
\begin{equation}
		X = -3e^{-3N/2}\int^N_{\tilde N_i} d\bar Ne^{3\bar N/2} = -2 + 2e^{-3(N - \tilde N_i)/2}\;,
\end{equation}
where $\tilde N_i$ is some new initial value, chosen in the matter-dominated epoch (say e.g. $\tilde N_i = -3$). Being $N - \tilde N_i$ always positive, the exponential part of the above solution rapidly becomes negligible with respect to $-2$ and so a linear solution for $U$ follows:
\begin{equation}
	U = C_4 - 2N\;,
\end{equation}
with $C_4$ another integration constant. This also is qualitatively successful, as seen in Fig.~\ref{Fig:hU}. Moreover, for $Y$ we have: 
\begin{equation}
	Y(N) = -e^{-3(N-N_i)/2}\int^N_{\tilde N_i} d\bar Ne^{3(\bar N-\tilde N_i)}\frac{m^2}{H_0^2\Omega_{\rm m0}e^{-3\bar N}} = -\frac{2m^2}{9H_0^2\Omega_{\rm m0}}\left(e^{3N} - e^{9\tilde N_i/2 - 3N/2}\right)\;,
\end{equation}
from which:
\begin{equation}
	\tilde V = -\frac{2m^2}{90H_0^2\Omega_{\rm m0}}e^{3N} - \frac{4m^2}{27H_0^2\Omega_{\rm m0}}e^{9\tilde N_i/2 - 3N/2} + C_5\;.
\end{equation}
Assuming $\tilde V(\tilde N_i) = 1$, we then have:
\begin{equation}
	\tilde V = -\frac{2m^2}{27H_0^2\Omega_{\rm m0}}e^{3N} - \frac{4m^2}{27H_0^2\Omega_{\rm m0}}e^{9\tilde N_i/2 - 3N/2} + \frac{6m^2}{27H_0^2\Omega_{\rm m0}}e^{4\tilde N_i} + 1\;.
\end{equation}
Neglecting all the exponentials contributions containing $\tilde N_i$ we can write:
\begin{equation}
	\tilde V = 1 - \frac{2m^2}{27H_0^2\Omega_{\rm m0}}e^{3N}\;.
\end{equation}
We can then conclude that it is only at late times, after the matter-dominated case, that $\tilde V$ starts to grow different from one.

From the Friedmann equation \eqref{Friedeq}, we have that:
\begin{equation}\label{modFriedeq3}
	3\tilde V = -\frac{m^2U}{2H^2} - \frac{Y}{2}(6 + X) + \frac{\rho}{M_{\rm Pl}^2H^2}\;.
\end{equation}
Since $U < 0$, $Y < 0$, $X > -6$ and of course $\rho > 0$, we can conclude that $\tilde V > 0$. On the other hand, $Y < 0$ tells us that $\tilde V$ always decreases. So, in order for $\tilde V$ to decrease from one to zero, without becoming negative, we need that $m^2/H^2 \ge 1$ only for a limited interval of e-folds. This, in particular, means that $H$ cannot tend to zero in the far future, for large $N$, but it must increase in order to guarantee that $m^2/H^2 \ll 1$. On the basis of this argument, we can conclude that at late-times, when matter is completely diluted, one has:
\begin{equation}
	3\tilde V \sim -\frac{m^2U}{2H^2}\;.
\end{equation}
This result however holds true provided that $X$ does not diverge, otherwise we cannot neglect the product $XY$ in Eq.~\eqref{modFriedeq3}.

Combining the two Friedmann equations \eqref{Friedeq} and \eqref{acceq}, we have that:
\begin{equation}
	\xi = -3 + \frac{1}{\tilde V}\left[-Y + \frac{m^2(1 - U)}{2H^2} + \frac{\rho - P}{2M_{\rm Pl}^2H^2}\right]\;.
\end{equation}
For non-exotic fluids one has $\rho - P > 0$ and therefore we see that definitely $\xi > -3$. At late times, according to our previous discussion, we have that:
\begin{equation}
	\xi \sim -3 - \frac{1}{\tilde V}\frac{m^2U}{2H^2} \sim 0\;.
\end{equation}
Hence, the effective equation of state always tends to $-1$, which is the main result of the present section. We can check now directly that $X$ does not diverge from its solution, computed with $\xi = 0$. One obtains from Eq.~\eqref{Xsol} for $\xi = 0$:
\begin{equation}
	X = -4\left[1 - e^{-3(N - \tilde N_i)}\right]\;,
\end{equation}
where $\tilde N_i$ is some initial e-fold number already within the epoch in which $\xi = 0$. We neglect this contribution and thus one can write the following solution for $U$:
\begin{equation}
	U = C_6 - 4N\;,
\end{equation} 
with another integration constant $C_6$. Again, this solution is seen to be in very good qualitative agreement with the numerical one in Fig.~\ref{Fig:hU}. With this solution, we are left with two equations:
\begin{eqnarray}
Y + 3\tilde V = -\frac{m^2}{2H^2}U\;,\qquad Y' + 3Y = -\frac{m^2}{H^2}\;.
\end{eqnarray}
The derivative of the left hand side of the first equation is equal to the left hand side of the second equation. Hence, one finds that:
\begin{equation}
	\xi = -\frac{3}{C_6 - 4N} = -\frac{3}{U}\;,
\end{equation}
and we have the solution for $H$:
\begin{equation}
	H = C_7|C_6 - 4N|^{3/4} = 3|U|^{3/4}\;,
\end{equation}
which means that $H$ keeps growing forever, never attaining a constant value, since in fact there are no de Sitter attractors, as we proved in Sec.~\ref{Sec:nonlocalmodel}. The solution for $H'$ is instead:
\begin{equation}
	H' = -\frac{3C_7}{C_6 - 4N}|C_6 - 4N|^{3/4}\;.
\end{equation}
In summary, the main finding of this section is a proof that the effective equation of state always tends to $-1$ or, equivalently, that $\xi$ always tends to zero. We have also provided analytic solutions for $H$, $U$ and $V$ in very good agreement with numerical calculations.

In the next section we address perturbations of the nonlocal model and how nonlocality manifests itself on Solar System scales and in the growth of small cosmological perturbations on small scales.

\section{Newtonian limit}\label{Sec:Newtonian Limit}

As shown in Fig.~\ref{Fig:hU}, Eqs.~\eqref{Friedeq} and \eqref{acceq} allow a viable cosmological history compatible with the $\Lambda$CDM model for a suitable choice of the free parameter $m^2$. In this section we investigate how nonlocality manifests itself on small scales.

\subsection{First-order scalar perturbations} 

We adopt the same procedure of Ref.~\cite{Koivisto:2008dh}, but we consider only scalar perturbations of the FLRW metric in the Newtonian gauge:
\begin{equation}
    ds^2 = -dt^2\left(1 + 2\psi\right) + a^2\left(1 + 2\phi\right)\delta_{ij}dx^idx^j\;,
\end{equation}
with as usual $a(t)$ function of the cosmic time only and $\psi(\mathbf x,t)$ and $\phi(\mathbf x,t)$ the gravitational potentials, functions of space and time. In the same way, we also split the localised fields $U$ and $V$ in a background plus perturbed part, i.e.
\begin{equation}
	U(\mathbf x,t) = U_0(t) + \delta U(\mathbf x,t)\;, \qquad V(\mathbf x,t) = V_0(t) + \delta V(\mathbf x,t)\;,
\end{equation}
where the background contributions depend only on the time. We omit the explicit functional dependences from now on.

The perturbed Einstein equations can be written as follows:
\begin{eqnarray}
	\left(1 - 2\alpha V_0\right)\delta G^\mu{}_\nu - 2\alpha\delta VG^{(0)\mu}{}_\nu - \frac{m^2\delta U}{2}\delta^\mu{}_\nu\nonumber\\ 
	+ 2\alpha\left[g^{(0)\mu\rho}\partial_\nu\partial_\rho\delta V + \delta g^{\mu\rho}\partial_\nu\partial_\rho V_0 - g^{(0)\mu\rho}\Gamma^{(0)\sigma}_{\nu\rho}\partial_\sigma\delta V - g^{(0)\mu\rho}\delta\Gamma^{\sigma}_{\nu\rho}\partial_\sigma V_0 - \delta g^{\mu\rho}\Gamma^{(0)\sigma}_{\nu\rho}\partial_\sigma V_0\right]\nonumber\\
	+ \alpha\delta^\mu{}_\nu\left(g^{(0)\rho\sigma}\partial_\sigma V_0\partial_\rho\delta U + g^{(0)\rho\sigma}\partial_\sigma\delta V\partial_\rho U_0 + \delta g^{\rho\sigma}\partial_\sigma V_0\partial_\rho U_0\right)\nonumber\\
	- 2\alpha\left(g^{(0)\mu\rho}\partial_{(\rho}\delta U\partial_{\nu)}V_0 + g^{(0)\mu\rho}\partial_{(\rho}U_0\partial_{\nu)}\delta V + \delta g^{\mu\rho}\partial_{(\rho}U_0\partial_{\nu)}V_0\right) = \frac{\delta T^{\mu}{}_\nu}{M_{\rm Pl}^2}\;.\qquad
\end{eqnarray}
Since the background localised fields depend only on time, we can simplify the above expression as follows:
\begin{eqnarray}
	\left(1 - 2\alpha V_0\right)\delta G^\mu{}_\nu - 2\alpha\delta VG^{(0)\mu}{}_\nu - \frac{m^2\delta U}{2}\delta^\mu{}_\nu\nonumber\\ 
	+ 2\alpha\left[g^{(0)\mu\rho}\partial_\nu\partial_\rho\delta V + \delta g^{\mu\rho}\partial_\nu\partial_\rho V_0 - g^{(0)\mu\rho}\Gamma^{(0)\sigma}_{\nu\rho}\partial_\sigma\delta V - g^{(0)\mu\rho}\delta\Gamma^{0}_{\nu\rho}\dot V_0 - \delta g^{\mu\rho}\Gamma^{(0)0}_{\nu\rho}\dot V_0\right]\nonumber\\
	+ \alpha\delta^\mu{}_\nu\left(-\dot V_0\dot{\delta U} - \dot{\delta V}\dot U_0 + 2\psi\dot V_0\dot U_0\right)\nonumber\\
	- 2\alpha\left(g^{(0)\mu\rho}\partial_{(\rho}\delta U\partial_{\nu)}V_0 + g^{(0)\mu\rho}\partial_{(\rho}U_0\partial_{\nu)}\delta V + \delta g^{\mu\rho}\partial_{(\rho}U_0\partial_{\nu)}V_0\right) = \frac{\delta T^{\mu}{}_\nu}{M_{\rm Pl}^2}\;.\qquad
\end{eqnarray}
The $0-0$ component of the perturbed Einstein equations is then:
\begin{eqnarray}
	\left(1 -2\alpha V_0\right)\delta G^0{}_{0} - 2\alpha\delta VG^{(0)0}{}_{0} - m^2\frac{\delta U}{2} \nonumber\\+ 2\alpha\left(-\ddot{\delta V} + 2\psi\ddot V_0 + \dot\psi\dot V_0\right) + \alpha\left(\dot{\delta U}\dot V_0 + \dot U_0\dot{\delta V}\right) - 2\alpha\psi \dot U_0\dot V_0 = -\frac{\delta\rho}{M_{\rm Pl}^2}\;.
\end{eqnarray}
with:
\begin{eqnarray}
	\delta G^0{}_{0} = -6H\dot\phi + 6H^2\psi + 2\frac{\nabla^2\phi}{a^2}\;, \qquad G^{(0)0}{}_{0} = -3H^2\;.
\end{eqnarray}
The $i-j$ component is:
\begin{eqnarray}
	\left(1 -2\alpha V_0\right)\delta G^i{}_{j} - 2\alpha\delta VG^{(0)i}{}_{j} - m^2\frac{\delta U}{2}\delta^i{}_{j} \nonumber\\ + \frac{2\alpha}{a^2}\partial^i\partial_j\delta V + 2\alpha\left(-H\dot{\delta V} + 2H\psi\dot V_0 - \dot\phi\dot V_0\right)\delta^i{}_{j}\nonumber\\ -\alpha\left(\dot{\delta U}\dot V_0 + \dot U_0\dot{\delta V} - 2\psi\dot U_0\dot V_0\right)\delta^i{}_{j} = \frac{\delta T^i{}_{j}}{M_{\rm Pl}^2}\;,
\end{eqnarray}
with
\begin{eqnarray}
	\delta G^i{}_{j} = \left[-2\ddot\phi - 6H\dot\phi + 2H\dot\psi + 4\dot H\psi - 2H^2\psi + \frac{1}{a^2}\nabla^2(\psi + \phi)\right]\delta^i{}_j - \frac{1}{a^2}\partial^i\partial_j(\phi + \psi)\;,\\ 
	G^{(0)i}{}_{j} = (-2\dot H - 3H^2)\delta^i{}_j\;.
\end{eqnarray}
For completeness, we include the $0-i$ equation, though we are not going to use it in the forthcoming analysis:
\begin{eqnarray}
	\left(1 -2\alpha V_0\right)\delta G^0{}_{i} - 2\alpha\delta VG^{(0)0}{}_{i} + 2\alpha\left(-\partial_i\dot{\delta V} + H\partial_i\delta V + \partial_i\psi\dot V_0\right)\nonumber\\ + \alpha\left(\dot V_0\partial_i\delta U + \dot U_0\partial_i\delta V\right) = \frac{\delta T^0{}_{i}}{M_{\rm Pl}^2}\;,
\end{eqnarray}
with
\begin{eqnarray}
	\delta G^0{}_{i} = 2\partial_i(\dot\phi - H\psi)\;, \qquad G^{(0)0}{}_{i} = 0\;.
\end{eqnarray}
Finally the evolution equations for the auxiliary fields are obtain perturbing the box operator:
\begin{eqnarray}
	\Box V = \Box^{(0)}V_0 + \Box^{(0)}\delta V + (\delta\Box)V_0\;,
\end{eqnarray}
where
\begin{eqnarray}
	\Box^{(0)}\delta V = g^{(0)\mu\nu}\partial_\mu\partial_\nu\delta V - g^{(0)\mu\nu}\Gamma^{(0)\rho}_{\mu\nu}\partial_\rho\delta V\;,\\
	(\delta\Box)V_0 = \delta g^{\mu\nu}\partial_\mu\partial_\nu V_0 - g^{(0)\mu\nu}\delta\Gamma^{\rho}_{\mu\nu}\partial_\rho V_0 - \delta g^{\mu\nu}\Gamma^{(0)\rho}_{\mu\nu}\partial_\rho V_0\;.
\end{eqnarray}
Of course, the same expressions hold true for $U_0$ and $\delta U$ as well. We can then write:
\begin{eqnarray}
	\Box^{(0)}\delta V = -\ddot{\delta V} + \frac{1}{a^2}\nabla^2\delta V - 3H\dot{\delta V}\;,\\
	(\delta\Box)V_0 = 2\psi\ddot V_0 + \dot\psi\dot V_0 + 3(2H\psi - \dot\phi)\dot V_0\;.
\end{eqnarray}
Finally, the perturbed Ricci scalar is, which enters the equation for $U$, is:
\begin{eqnarray}
	\delta R = -\frac{2}{a^2}\nabla^2(2\phi + \psi) - 6H\dot\psi - 12(\dot H + 2H^2)\psi + 6\ddot\phi + 18H\dot\phi\;. 
\end{eqnarray}

\subsection{Solar system scales}

In order to understand how the above equations describe gravity on solar system scales we make the following approximations, as in Ref.~\cite{Koivisto:2008dh}:
\begin{enumerate}
    \item We ignore the cosmological expansion, so we set the scale factor $a = 1$ and the Hubble factor $H = 0$.
    \item We look for a static solution for the gravitational potentials.
    \item We set matter perturbations to zero.
\end{enumerate}
When perturbing, we must remember that the model under investigation contemplates another metric $f_{\mu\nu}$ which enters the field equations through the Ricci scalar $R_f$, contained in $m^2 = -2\alpha R_f$. Therefore, the perturbations $\delta V$ couples to $\delta R_f$, as one can see from Eq.~\eqref{auxiliaryeqsNL}. On the other hand, the total $R_f$, i.e. background plus perturbation, must be constant because of the Bianchi constraints and hence $\delta R_f$ must also be constant.

Now we come to a crucial consideration. When we set $a = 1$ and consider vacuum we have Minkowski space as background. Let us write Eqs.~\eqref{modFriedeq} and \eqref{modacceq} and \eqref{UandVeqs} for Minkowski space:
\begin{eqnarray}
	-m^2(1 - U_0/2) + m^2 - \alpha\dot U_0\dot V_0 = 0\;, \qquad m^2(1 - U_0/2) - \alpha\dot U_0\dot V_0 = 0\;,\\
	\ddot U_0 = 0\;, \qquad \ddot V_0 = \frac{m^2}{2\alpha}\;.
\end{eqnarray}
These equations demand that $m^2 = 0$, i.e. it is not possible to have Minkowski space as solution if $R_f$ is not vanishing. So, we now consider $m^2$ to be a perturbative quantity (we want to avoid writing $\delta(m^2)$). In order to satisfy the above equations when $m^2 = 0$ we just need either $\dot U_0 = 0$ or $\dot V_0 = 0$. To make it simple, and natural, we choose $U_0 = V_0 = 0$.

Therefore, we can cast the equation for $\delta U$ as follows, using only $H = 0$ and $U_0 = V_0 = 0$:
\begin{equation}
	-\ddot{\delta U} + \nabla^2\delta U = -2\nabla^2(2\phi + \psi)\;.
\end{equation}
Since we look for static gravitational potentials, we must have a time-independent $\delta U$. We are then left with:
\begin{equation}
	\nabla^2\delta U = -2\nabla^2(2\phi + \psi)\;.
\end{equation}
The equation for $\delta V$, again only setting $H = 0$ and $U_0 = V_0 = 0$:
\begin{equation}
	-\ddot{\delta V} + \nabla^2\delta V = -\frac{m^2}{2\alpha}\;.
\end{equation}
From the field equations we will later notice that we need $\delta V$ to be time-independent in order to have static potentials, hence:
\begin{equation}
	\nabla^2\delta V = -\frac{m^2}{2\alpha}\;,
\end{equation}
where recall that $m^2 = -2\alpha\delta R_f$. For the perturbed $0-0$ modified Einstein equations we have:
\begin{eqnarray}
	\nabla^2\phi = 0\;. 
\end{eqnarray}
since, according to our analysis, $m^2\delta U$ is a second order quantity. The $i-j$ component is:
\begin{eqnarray}
	\nabla^2(\psi + \phi)\delta^i{}_j - \partial^i\partial_j(\phi + \psi) + 2\alpha\partial^i\partial_j\delta V = 0\;,
\end{eqnarray}
again neglecting $m^2\delta U$ as a second order quantity. The $0-i$ equation is identically vanishing. In summary, we have the following set of equations:
\begin{eqnarray}
	\nabla^2\phi = 0\;,\\
	\nabla^2(\psi + \phi) + \alpha\nabla^2\delta V = 0\;,\\
	-(\partial^i\partial_j - \delta^i{}_j\nabla^2/3)(\phi + \psi - 2\alpha\delta V) = 0\;,\\
	2\alpha\nabla^2\delta V = -m^2\;,\\
	\nabla^2\delta U = -2\nabla^2(2\phi + \psi)\;.
\end{eqnarray}
The first equation is the standard Poisson equation (or Laplace equation, since we are in vacuum), which nonlocality leaves unaffected. Assuming spherical symmetry, the solution is:
\begin{equation}
	\phi = \frac{C_1}{r} + C_2\;,
\end{equation} 
which can be recast, by demanding that the potential vanishes at infinity and that we recover Schwarzschild solution (in isotropic coordinates) in the GR limit (which corresponds to $m^2 = 0$) as:
\begin{equation}
	\phi = \frac{GM}{r}\;.
\end{equation}
The gravitational potential $\psi$ is instead given as solution of the following equation:
\begin{equation}
	\nabla^2\psi = \frac{m^2}{2} = -\alpha\delta R_f\;,
\end{equation}
i.e. it is sourced by the perturbed secondary Ricci scalar. We get then, again assuming spherical symmetry:
\begin{equation}
	\psi = \frac{m^2r^2}{12} - \frac{GM}{r}\;,
\end{equation}
where we have chosen the integration constants in order to recover Schwarzschild solution for $m^2 = 0$. The first term $m^2r^2$ is quite interesting because it is how we expect a cosmological constant to modify the gravitational potential on local scales, if $m^2 > 0$.

Given the explicit expressions of the two gravitational potentials, the post-Newtonian parameter $\gamma$ is thus:
\begin{equation}
	\gamma \equiv -\frac{\phi}{\psi} = \frac{1}{1 - \frac{m^2r^3}{12GM}}\;.
\end{equation}
Observational constraints, see e.g. Ref.~\cite{Will:2014kxa}, establish that $|\gamma - 1| \lesssim 10^{-5}$, hence we must have that:
\begin{equation}
	\frac{m^2r^3}{12GM} \lesssim 10^{-5}\;.
\end{equation}
Since $m^2$ must be constant for the internal consistency of the theory and $r$ can be very large (even of the order of galactic scales and beyond, depending on the test performed to determine $\gamma$) one can see that $m^2$ has to be vanishingly small. This is not however a concern, since the nonlocal modification of GR is devised in order to be relevant on cosmological scales, hence $m^2 \sim H_0^2$.

\subsection{Structure formation}

We consider in the present instance pressureless dust as the only component. Combining its continuity and Euler equations, see e.g.  Ref.~\cite{Piattella:2018hvi}, we obtain:
\begin{equation}\label{matterequation}
\ddot{\delta} + 2H\dot{\delta} + 3\ddot{\phi} + 6H\dot{\phi} + \frac{k^2}{a^2}\psi = 0 \;,    
\end{equation}
where $\delta \equiv \delta\rho/\rho$ is the density contrast of dust and we have also introduced the Fourier transform, being $k$ the comoving wavenumber. Following Ref.~\cite{Koivisto:2008dh} we consider a quasi-static approximation (QSA), i.e. we consider scales such that $k^2 \gg H^2$, thereby we neglect all the time derivatives and the terms proportional to $H$ and its derivative with respect to $k^2$ for all the perturbative quantities. Moreover, in this instance we consider $\delta R_f = 0$, since the background FLRW solution is consistent with a $m^2 \neq 0$.  Under these assumptions, the equation for $\delta$ can be simplified as follows:
\begin{equation}
\ddot{\delta} + 2H\dot{\delta} + \frac{k^2}{a^2}\psi = 0 \;,    
\end{equation}
and the equations for $\delta U$ and $\delta V$ become:
\begin{eqnarray}
	\delta U = -2\left(\psi + 2\phi\right)\;, \qquad 	\frac{k^2}{a^2}\delta V = \frac{m^2}{\alpha}\psi\;.  
\end{eqnarray} 
The modified Poisson equation and the spatial trace of the field equations become:
\begin{eqnarray}
	2\left(1 - 2\alpha V_0\right)\frac{k^2\phi}{a^2} + m^2\frac{\delta U}{2} - 2m^2\psi = \frac{\delta\rho}{M_{\rm Pl}^2}\;,\\
	-2\left(1 -2\alpha V_0\right)\frac{k^2}{a^2}(\psi + \phi) - 3m^2\frac{\delta U}{2} - 2\alpha\frac{k^2}{a^2}\delta V = 0\;,
\end{eqnarray}
where recall that the pressure perturbation for dust is also zero ad we are assuming that pressureless dust as the only component. The second equation establishes a relation between the gravitational potentials, using also the equations for $\delta U$ and $\delta V$:
\begin{equation}
	-\left[2\left(1 -2\alpha V_0\right)\frac{k^2}{a^2} - 6m^2\right]\phi = \left[2\left(1 -2\alpha V_0\right)\frac{k^2}{a^2} - m^2\right]\psi\;,
\end{equation}
from which we infer the following slip parameter:
\begin{equation}
	\eta \equiv -\frac{\phi}{\psi} = \frac{2\left(1 -2\alpha V_0\right)k^2 - m^2a^2}{2\left(1 -2\alpha V_0\right)k^2 - 6m^2a^2}\;.
\end{equation}
The GR result $-\phi = \psi$ is obtained for $m^2 = 0$, as expected. Substituting the above result in the modified Poisson equation, we obtain:
\begin{equation}
	\frac{\left(1 -2\alpha V_0\right)^2\frac{k^4}{a^4} - 4m^4}{\left(1 -2\alpha V_0\right)\frac{k^2}{a^2} - 3m^2}\psi = -\frac{\delta\rho}{2M_{\rm Pl}^2}\;,
\end{equation}
and hence the equation for $\delta \equiv \delta\rho/\rho$ becomes:
\begin{equation}
\ddot{\delta} + 2H\dot{\delta} - \frac{\left(1 -2\alpha V_0\right)k^4 - 3m^2k^2a^2}{\left(1 -2\alpha V_0\right)^2k^4 - 4m^4a^4}\frac{3}{2}H^2\delta = 0\;.
\end{equation}
Using $N = \ln a$ as independent variable, we get:
\begin{equation}
\delta'' + (2 + \xi)\delta' - \frac{\left(1 -2\alpha V_0\right)k^4 - 3m^2k^2a^2}{\left(1 -2\alpha V_0\right)^2k^4 - 4m^4a^4}\frac{3}{2}\delta = 0\;.    
\end{equation}
The effective gravitational coupling is then:
\begin{equation}
	Y = \frac{\left(1 -2\alpha V_0\right)k^4 - 3m^2k^2a^2}{\left(1 -2\alpha V_0\right)^2k^4 - 4m^4a^4}\;.
\end{equation}
As one can see from the above analysis, nonlocality, embodied in the parameter $m^2$, affects gravity on local scales as we would expect a cosmological constant to do so. On cosmological scales, which are much smaller than the Hubble radius though, nonlocality manifests itself as a modification of the slip parameter and the effective gravitational coupling. Since for a viable background expansion we have $m^2 \sim H_0^2$, in the QSA approximation we have then that:
\begin{equation}
	\eta \approx 1\;, \qquad Y \approx \frac{1}{1 - 2\alpha V_0}\;,
\end{equation} 
i.e. the slip parameter is essentially unity, the same as in GR, whereas the effective gravitational coupling gets larger as soon as nonlocality kicks in driving the accelerated expansion of the universe, since $1 - 2\alpha V_0$ becomes smaller than unity, as it can be seen from Fig.~\ref{Fig:UV}. For example, using $m^2 = 0.2H_0^2$, we can infer from Fig.~\ref{Fig:UV} that $1 - 2\alpha V_0 = \tilde V_0 \approx 0.98$ and hence $Y \approx 1.02$, i.e. the gravitational coupling strength is enhanced by 2\%.

\section{Conclusion}\label{Sec:Concl}

In this paper we have investigated the features of the cosmological expansion history described by the non-local bi-metric interacting model of Ref.~\cite{Vardanyan:2017kal}. We have performed in Sec.~\ref{Sec:nonlocalmodel} a detailed analysis of the dynamical system~\eqref{dsfinitedistance2} formed by the field equations and we have found no stable critical points at finite distance. We have also looked for critical points at infinite distance, studying the dynamical system \eqref{dsinfinitedistance1} and we have also found no stable critical points at infinite distance.

The absence of a de Sitter attractor might rise the doubt that any phase of accelerated expansion should be transitory. In order to establish if this is the case, in Sec.~\ref{Sec:Qualitativeanalysis} we have solved numerically the field equations \eqref{systemVAAS} in order to capture how nonlocality affects the evolution of the Hubble parameter and of the localised fields $U$ and $V$. We have showed that the effective equation of state $\omega_{\rm eff}$ always tends to $-1$, independently from the value of the free parameter $m^2$. This means that the expansion is forever accelerated, as it happens in the $\Lambda$CDM model, even if the Hubble parameter does not tend to a constant but it grows instead. It is worth noting that such a feature is shared also by the Maggiore-Mancarella model \cite{Maggiore:2014sia}, as was showed in Ref.~\cite{Nersisyan:2016hjh}. It is an interesting issue to establish whether this is a coincidence or not, but we leave this investigation for a future work.

Finally, in Sec.~\ref{Sec:Newtonian Limit} we have considered first order scalar perturbations on flat FLRW background in order to study the Newtonian limit of the model and how small fluctuations on small scales, where the quasi-static approximation is applicable, grow. We have shown that nonlocality induces a post-Newtonian parameter $\gamma$, a gravitational slip and an effective normalised gravitational coupling which are different from unity and therefore different from those in GR. These corrections are nonetheless negligibly small if one assumes for $m^2$ the value necessary to reproduce a viable cosmological history, i.e. $m^2 \sim H_0^2$, thereby making the model viable in the aforementioned regimes. A constraint on $m^2$ based on the most precise cosmological observations and a more comprehensive perturbative analysis of the model are still lacking and they will be addressed in a forthcoming paper. 

\subsection*{Acknowledgements}

The authors are grateful to Luca Amendola for useful comments and suggestions.  

This study was financed in part by the \emph{Coordena\c{c}\~ao de Aperfei\c{c}oamento de Pessoal de N\'ivel Superior} - Brazil (CAPES) - Finance Code 001. OFP thanks the Alexander von Humboldt foundation for funding and the Institute for Theoretical Physics of Heidelberg University for kind hospitality.  

\bibliographystyle{unsrturl}
\bibliography{NL-DS}

\end{document}